\documentclass[twocolumn,floatfix, showpacs]{revtex4}

\usepackage[final]{epsfig}
\usepackage{amsmath}
\usepackage{parskip}

\begin{document}

\title{Three-particle azimuthal correlations and Mach shocks}

\author{Thorsten Renk}


\affiliation{Department of Physics, PO Box 35, FIN-40014 University of Jyv\"askyl\"a, Finland}
\affiliation{Helsinki Institute of Physics, PO Box 64, FIN-00014, University of Helsinki, Finland}
\author{J\"org Ruppert}
\affiliation{Department of Physics, McGill University, 3600 Rue University, Montreal, Quebec, Canada, H3A 2T8}

\begin{abstract}
Measurements of angular correlations of hadrons with a (semi-)hard trigger
hadron in heavy-ion collisions at RHIC show large angular structures
opposite to the trigger which were a priori unexpected. 
These away side large angle correlations were first observed in two-particle
correlations \cite{PHENIX-2pc,Adams:2005ph} and have recently also been investigated in three-particle
correlation measurements \cite{Ulery:2005cc,Wang:2006ig}.
We show that the correlation signal can be understood in terms of sonic shockwaves ('Mach cones') excited by hard partons supersonically traversing the medium.
The propagation of such shocks through the medium evolution is treated in a Monte Carlo (MC) framework \cite{Renk:2005si,Renk:2006mv}. We demonstrate that two- and especially three-particle correlations offer non-trivial insight into the medium-averaged speed of sound and the evolution of flow.  Our findings imply that the assumption of "deflected jets" is not necessary to account for the observed correlations. 
\end{abstract}

\maketitle

% main text

The four detector collaborations at RHIC have announced that a new state 
of matter distinct from ordinary nuclear matter has been observed in ultra-relativistic heavy-ion collisions \cite{RHIC-QGP}. The interaction of hard partons created at the first moments of the collision with the soft bulk medium has been suggested as a promising probe to study the properties of this state of matter \cite{Jet1,Jet2,Jet3,Jet4,Jet5,Jet6}. 
Recent measurements of two- and three particle correlations involving a semi-hard trigger parton with $2.5~{\rm GeV}< p_T< 4.0~{\rm GeV}$ and an associate threshold of $1.0~{\rm GeV}<p_T<2.5~{\rm GeV}$ have shown an a priori surprising splitting of the away-side peak for all centralities but peripheral collisions which is very different from the broadened away-side peaks observed in p-p and d-Au collisions \cite{PHENIX-2pc}.
The structure of the away-side peak is also persistent for different associate hadron momentum cuts (cf. \cite{Adams:2005ph}). 
  
The observed structures have been interpreted in terms of interactions of the away-side
parton with the soft medium. In particular the excitation of colorless and colored collective modes has been suggested, see \cite{Stoecker,Casalderrey-Solana:2004qm,Wake,Koch:2005sx}.  The predictions inferred from the excitation of sound modes  \footnote{An alternative explanation in terms of  Cherenkov radiation predicts a  shrinking of the emission angle with increasing momentum of the associate hadrons \cite{Koch:2005sx} which is not observed. The modification of the space-like gluon dispersion relation in this scenario is calculated in a bound state model \cite{Koch:2005sx}. Note also our discussion in \cite{Renk:2006mv}.} turn out to be in good agreement with two-particle correlation data \cite{Renk:2005si,Renk:2006mv}.  
Model calculations for Mach shocks in a true hydrodynamical framework have so far only been performed using a boost-invariant 'Mach-wedge' as an approximation of the situation close to midrapidity \cite{Casalderrey-Solana:2004qm,Chaudhuri:2005vc}.  However, this is not the experimental situation in which the trigger condition only constraints the near-side hadron to be at rapidity $y=0$. The position of the away-side parton  can be inferred as a  conditional probability distribution $P(y)$ calculable in perturbative QCD. We have shown in \cite{Renk:2006mv} that the measured correlation structure arises from folding of $P(y)$ with the rapidity structure of the correlation for single away-side
partons, taking detector acceptance into account. However, as the correlation signal from single partons is elongated in $\phi,y$-space by longitudinal flow this does not lead to a strong reduction of the predicted maximum correlation angle which would be in contradiction to experiment, see \cite{Renk:2006mv} and especially Fig. 3 therein.

\begin{figure}[htb]
\vspace{0.5cm}
\epsfig{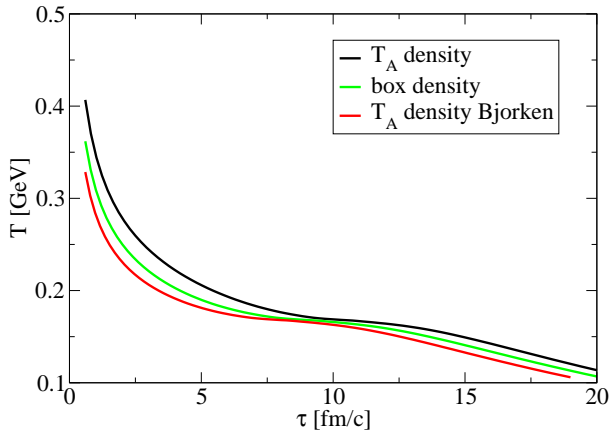}
\caption{\label{F-1} Core temperature evolution in 200 AGeV Au-Au collisions from fireball models with three
different transverse entropy density profiles and accelerated longitudinal or boost-invariant longitudinal expansion respectively (see text).}
\end{figure}

Here we focus on central collisions. In order to calculate two- and three-particle correlations we proceed as follows (for details see \cite{Renk:2005si,Renk:2006mv}): We simulate the trigger condition as closely as possible by using a MC approach. The hard vertices are generated with a distribution weighted by the nuclear overlap  and parton type and momenta are determined by randomly sampling partonic transverse momentum spectra as generated by the VNI/BMS parton cascade as described in \cite{Renk:2005yg}. For the energy loss calculation we use the formalism outlined in Ref. \cite{QuenchingWeights} to obtain the probability $P(\Delta E)$ of the parton to lose the amount of energy $\Delta E$
from two characteristic quantities, namely the plasma frequency $\omega_c$
%\begin{eqnarray}
 %\omega_c({\bf r_0},\phi)=\int_0^\tau d \xi \xi \hat{q}(\xi)
%\end{eqnarray}
and the averaged momentum transfer $\langle\hat{q} L\rangle({\bf r_0}, \phi)$.
%\begin{eqnarray}
%({\hat{q}L}({\bf r_0}, \phi))=\int_0^\tau d \xi {\hat{q}(\xi)}
%\end{eqnarray}
%in a static scenario calculated along the path of the hard parton through the medium.
For the description of the dynamical evolution the fireball model outlined in Refs. \cite{RenkSpectraHBT,Synopsis,Photons_RHIC1,Photons_RHIC2} is used. It is constrained by transverse mass spectra of hadrons, Hanbury-Brown-Twiss (HBT) correlation radii and hadronic $dN/dy$ distributions and is known to lead to a good description of direct photon yields as well as $R_{AA}$
in the high $p_T$-region, \cite{Renk:2005ta,Renk:2006sx}. It is also successful in predicting the back-to-back correlations for a hard trigger $>8$ GeV   \cite{Renk:2006pk}.
Here we study three different cases for the dynamics of the fireball. 
The first one is referred to as 'box profile' (the model described in Ref. \cite{RenkSpectraHBT}) and is characterized by a transverse Woods-Saxon entropy density profile with a small surface thickness $d_{\rm ws} \approx 0.5$ fm. This distribution is favoured by a simultaneous fit to transverse mass spectra and HBT correlation radii. The second utilizes the nuclear profile $T_A$ as transverse density distribution as expected for a soft bulk matter production mechanism. In the third case we retain the $T_A$ density profile but change the accelerated longitudinal expansion of \cite{RenkSpectraHBT} into a boost-invariant Bjorken evolution. The other geometrical scales of the model are as in the first case. Equal temperature contours for the three cases are discussed in \cite{Renk:2006pk} (Fig.~3 therein). The core temperature evolution as a function of proper time $\tau$ for central $200 {\rm AGeV}$ Au-Au collisions for the three different cases is shown in Fig. \ref{F-1}. 

\hspace{0.5cm}

We calculate the energy lost from the near-side parton and decide if the trigger condition is fulfilled. If the experiment triggers on a semi-hard hadron in the transition regime where both recombination and fragmentation might be important, only the parton is required to fulfill the trigger condition for simplicity. We have checked that the shape of the correlation signal as predicted by the model is not sensitive to the exact value of the trigger threshold (we see however an uncertainty in the absolute normalization).  If the near-side parton passes the trigger condition, the direction of the away-side parton is determined taking into account intrinsic $k_T$. In this way one finds for each scenario a specific distribution of initial vertices, directions and momenta of away-side partons, i.e. all trigger bias is taken into account consistently. In the MC, the away side correlation is inferred from a sufficiently large sample of generated events. For every single event the averaged away side energy loss per proper time  $dE/d\tau$ is calculated. Our main assumption is that a fraction $f$ of the lost energy excites a collective mode of the medium which is characterized by a dispersion relation $E=c_s p$, where the speed of sound $c_s$ is determined locally by the equation of state as fitted to lattice results \cite{STW}. This determines the initial angle of propagation of the shock front with the jet axis, $\phi= \arccos c_s$. Elements of the front are subsequently propagated with the local 
 $c_s$ and $\phi$ is continoulsy adjusted as
\begin{equation}
\phi=\arccos \frac{\int_{\tau_{\rm E}}^\tau d \tau' c_s(\tau')}{\tau-\tau_{\rm E}},
\end{equation}
where $c_s(\tau)$ is calculated along the path of propagation.

 \vspace{0.5cm}
\begin{figure}[htb]
\vspace{-0.5cm}
\epsfig{file=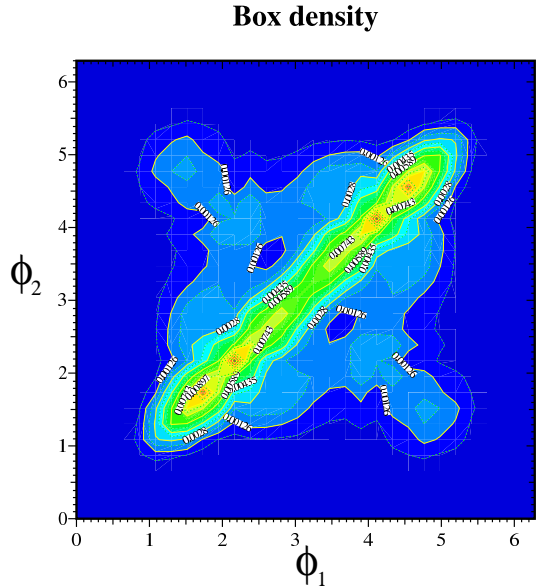, width=7cm}\\ 
\caption{\label{F-3} Calculated three-particle correlation signal as inferred from Mach shocks created by supersonically traveling partons in the medium on the away-side. A Monte Carlo simulation accounts for the near-side trigger condition.
The transverse entropy density is distributed according to a Wood-Saxon form with 0.5 fm surface thickness ('box density'). 
}
\end{figure}

\vspace{0.5cm}
\begin{figure}[htb]
\vspace{-0.5cm}
%l\epsfig{file=corr-3p-box.eps, width=7cm}\\ \vspace{-1.5cm}
\epsfig{file=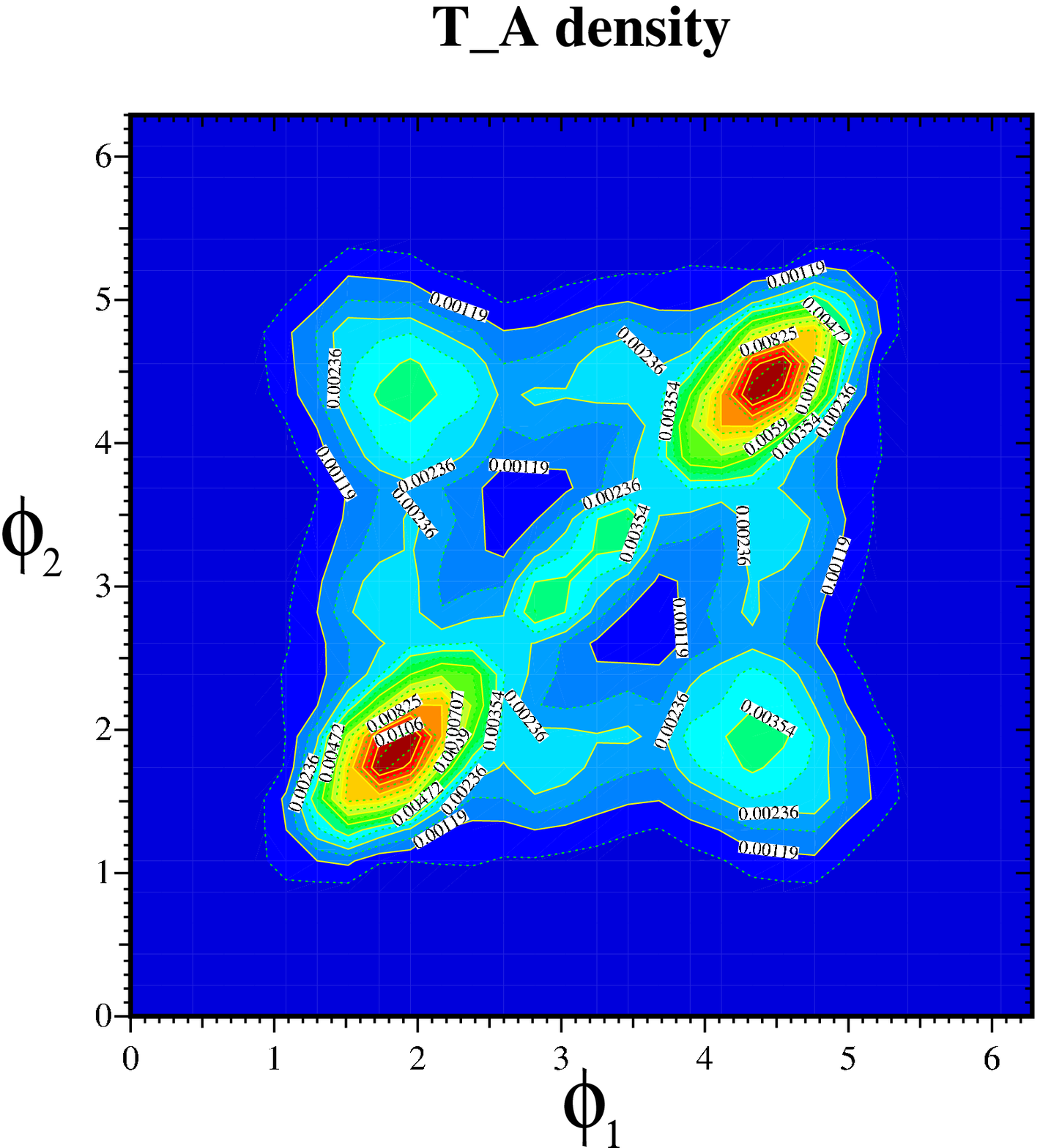,width=7cm}\\ 
\epsfig{file=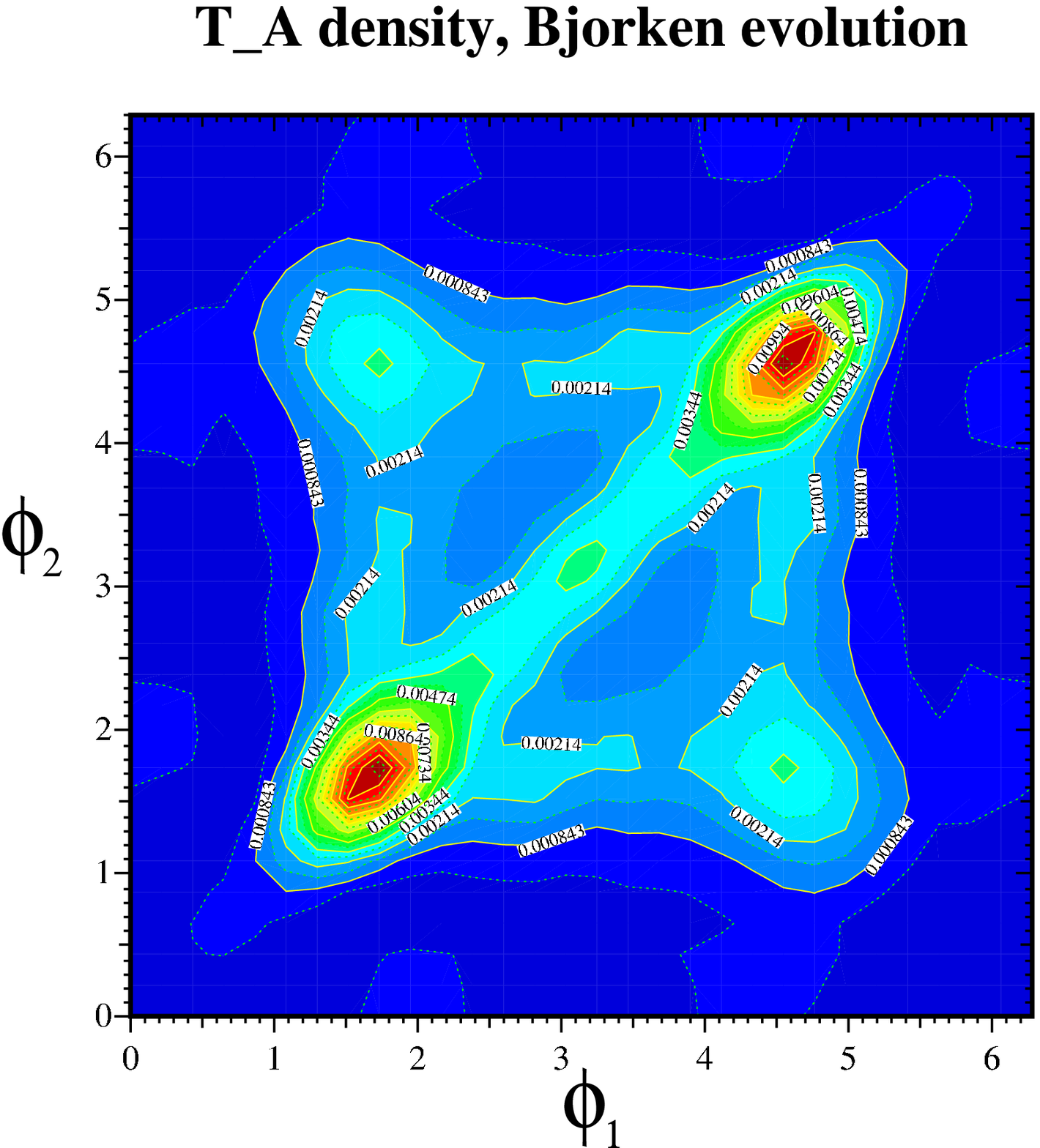,width=7cm}
\caption{\label{F-2} Calculated three-particle correlation signal as inferred from Mach shocks created by supersonically traveling partons in the medium on the away-side. A Monte Carlo simulation accounts for the near-side trigger condition. Upper panel:
transverse entropy density  distributed according to the nuclear profile $T_A$; accelatered longitudinal expansion. Lower panel: transverse density as above, but boost-invariant longitudinal expansion.}
\end{figure}

 Since the mode is propagated with $c_s$ relative to the surrounding medium, the shock front can be significantly distorted in {\it position} space by local transverse 
and  longitudinal flow \cite{Satarov:2005mv,Renk:2006mv}. Note that the final longitudinal {\it position} $z_{\rm final}$ at freeze-out is relevant only insofar as  the longitudinal flow at $z_{\rm final}$ determines a longitudinal boost in momentum space \cite{Renk:2006mv}. 
Once an element of the shock front reaches the freeze-out hypersurface, the sonic mode cannot propagate any further. We assume that the energy and momentum contained in the shock front are converted into additional kinetic energy and employ the matching condition to calculate the additional boost $u_{\rm shock}$ of a volume element received from the shock front. The Cooper-Frye formula is used  to convert
the fluid element into a hadronic distribution:

\begin{equation}
E \frac{d^3N}{d^3p}=\frac{g}{(2\pi)^3} \int d\sigma_\mu p^{\mu} \exp \left[ \frac{p^{\mu}(
u_\mu^{\rm flow}+u_\mu^{\rm shock})-\mu_i}{T_f}\right] \, .
\end{equation}

As discussed in \cite{Renk:2006es} section $4$ the observed yield is very sensitive to flow if the associate hadron cut is set well above the bulk matter momentum scales at freeze-out ($\sim 3~T_f$) and comes from those parts of the shock front where the transverse flow is (approximately) aligned with the emitted particle momentum. In the presence of a shockwave the signal at 1 ${\rm GeV}$ is in fact about $9$ times larger if $u^{\rm shock}_\mu \parallel u^{\rm shock}_\mu$  than if $u^{\rm shock}_\mu \perp u^{\rm shock}_\mu$. 
Thus, distortions in position space by flow perpendicular to the wave propagation do not map into a sizeable angular distortion of in momentum space since flow and shock have to be (approximately) aligned in every case were substantial contributions to the yield are expected.
On the other hand momentum conservation dictates that even if shock and flow are not aligned the correlation signal cannot vanish. In this case a broader correlation structure is observed at a lower $p_T$. 

This is important for the understanding of 2- and 3-particle correlation signals:  If the direction and initial position of the away-side parton in a given MC event is just such that transverse flow and cone are aligned, the expected correlation signal is very different from the case where transverse flow and the cone are orthogonal to each other (cf. left panel of Fig. $2$ in \cite{Renk:2006es}). In the aligned case both wings of the Mach cone leave their trace in the correlation signal while in the case were transverse flow is orthogonal to one wing of the shock cone the correlation signal can be suppressed and particles are redistributed to lower associate $p_T$.
Note that the case were only one wing appears in the correlation signal of a specific away-side jet could be misinterpreted as a "deflected jet".
The calculated two-particle correlation signal which is compared with the experimental result emerges in the Monte Carlo approach as an average of the signal expected from individual events. Large event by event fluctuations of signal shape and strength are observed. 
The averaging reduces information: if one would study a subsample of MC events in which always only one wing of the cone is visible, after averaging the signal would be difficult to distinguish from a situation in which always both sides of the cone are seen with half strength. It is therefore clear that an {\it averaged} two-particle correlation measurement can in principle not distinguish the two situations.

However, more information is carried by 3-particle correlations. Both the PHENIX and STAR collaboration have presented (preliminary) 3-particle
correlation measurements \cite{Ulery:2005cc,Wang:2006ig}. We chose the way of presenting the correlation signal in the $\phi_{12}$ and $\phi_{13}$ plane as utilized by the STAR collaboration. Here $\phi_{12}$ indicates the angle between the  trigger hadron and one of the associated hadrons  and $\phi_{13}$ the angle between trigger and other associated hadron.

Our Monte Carlo approach to calculate the two-particle correlation signal can be straightforwardly extended to calculations of the three-particle correlation signal in the away-side hemisphere assuming that 3-particle correlations can be calculated from factorized 2-particle correlations (as we do not simulate the complete near side jet but only the leading hadron, we do not show correlation signals inside the trigger hemisphere). Within this assumption, we neglect true 3-particle correlations. Since the Mach cone is a collective phenomenon it can involve typically of the order of $20-50$ particles, depending on the specific event under consideration \cite{Renk:2005si}.  While all of them are strongly  correlated with the original away-side parton (and hence with the trigger) and thus correlated among each other because of the conic geometry, these effects are part of the factorized two-particle correlations. Additional correlations of particles inside the cone are weak and are only induced by momentum conservation $\sum p_L=p_{\rm away}$ and $\sum {\bf p_T}=0$ (see also \cite{Renk:2005si}).

We show the results for the three-particle correlation pattern on the away side in Fig. 2 and Fig. 3 for the three different scenarios discussed above. Note that we assume for this calculation that the correlation measurement is performed exactly at midrapidity. All cases clearly indicate a correlation signal along the diagonal in the $\phi_{12}$ and $\phi_{13}$ plane and variable strength along the off-diagonal. 
The maxima along the diagonal mirror the maxima in the two-particle correlation measurements. The opening angle of the Mach cone (i.e. the distance between the maxima) is larger whenever the averaged $c_s$ is reduced.
Since $c_s$ is lowest in the vicinity of the phase transition \cite{STW}, the opening angle is the largest for the $T_A$ entropy density profile in the Bjorken evolution case, a bit less in the box entropy density case and least in the $T_A$ entropy density case with accelerated longitudinal expansion, cf. the $T$ evolution in Fig.~1. 

Correlation strength along the diagonal is expected if two correlated particles on the away-side are typically picked up from the same wing of the cone, while the off-diagonal structures appear if they originate from opposite wings. 
The three particle correlation signal for the box density profile has most strength on the diagonal, indicating that events in which only one side of the cone is aligned with the transverse flow at the freeze-out hypersurface occur more frequently.
The off-diagonal correlation signal is more pronounced in the $T_A$ entropy density profiles. This demonstrates that while off-diagonal correlation signals are a strong indicator for the existence of the formation of Mach cones on the away-side, their strength is sensitive to the details of the underlying evolution. We emphasize that it is not necessary to invoke the notion of a "deflected jet" as an explanation for the diagonal correlation structures.

This analysis demonstrates that the tomographic capabilities of correlation measurements are significantly increased by three particle correlations \cite{Ulery:2005cc,Wang:2006ig}. They open 
the possibility  to test properties of the excitation of collective modes \cite{Stoecker,Casalderrey-Solana:2004qm,Wake}. Furthermore they may also provide  a deeper insight into  the speed of sound in the nuclear medium as well as into our theoretical understanding of the evolution of the flow field and the bulk matter density in ultrarelativistic heavy-ion collisions.

\begin{acknowledgments}

This work was supported by the Academy of Finland, Project 206024,  and by the Natural Sciences and Engineering Research Council of Canada.

\end{acknowledgments}

\end{document}